# Adiabatic based Algorithm for SAT:
# a comprehensive algorithmic description

Eric Bourreau
LIRMM, CNRS
Montpellier University,
France
eric.bourreau@lirmm.fr

Gérard Fleury
LIMOS, CNRS
Clermont-Auvergne University,
France
gerard.fleury@isima.fr

Philippe Lacomme
LIMOS, CNRS
Clermont-Auvergne University,
France
philippe.lacomme@isima.fr

**ABSTRACT**

This paper concerns quantum heuristics able to extend the domain of quantum computing, defining a promising way in the large number of well-known classical algorithms. Quantum approximate heuristics take advantage of alternation between a Hamiltonian defining the problem to solve and a mixing Hamiltonian. The adiabatic theorem initially defined in quantum physic allows to compute a solution for the Schrödinger equation, but the foundation of this methods requires strong skill in physics and mathematics. Our main objectives in this paper are at first to provide an algorithm-based presentation (as close as possible of the classical computer science community in operational research practice) of the adiabatic optimization and secondly to give a comprehensive resolution of the well-known SAT problem. This gives opportunities to provide a concise but explicit analysis of the adiabatic capability to define a new efficient operational research trend. Our experiments encompass numerical evaluations on both simulator and on real quantum computer provided by IBM. Numerical experiments on simulator have been achieved on both Qiskit and MyQLM.

## 1. Introduction

A wide majority of Operational Research problems can be defined by the minimization of an objective function with the satisfaction of problem dependent constraints. Resolution approaches are based on exact methods, heuristic or metaheuristics depending on the problem and on the computation time available to provide one solution. From a practical point of view, heuristic and metaheuristic are commonly used in large instances resolutions and they address the following key-points during the search space investigation:

1. generation of initial solutions by powerful constructive heuristics dedicated to the problem;
2. application of a local search to obtain local minima and by consequence favor convergence;
3. diversification mechanisms to avoid premature convergence and search space trap;
4. Indirect representation of solutions to relax constraints and alternate between several search spaces.

Adiabatic based quantum approaches can be seen as a new promising local search heuristic. It uses gradually decreasing quantum fluctuations to traverse the barriers in the energy landscape. In fact, minimization of the objective function does not exist in the variable search space but in the representation space as the combinatorial problem is described as an operator. Exploration of the "Operator Search Space" allows gradually to compute better cost function. More precisely Adiabatic Quantum Optimization (AQO) (Fahri, 2000) takes advantage of alternating between the cost function modeled by the Hamiltonian $H_p$ and a mixing Hamiltonian operator $H_D$. In this paper, our goal is to give a complete understanding of the origin of the adiabatic optimization and deliver to the reader all the necessary mathematical basis to understand the main idea behind the scenes. Starting from the Schrödinger equation, we will gradually explain how to use the solution of this equation as a guess for an optimization problem described as a Hamiltonian. Deriving from an easy to define initial state, we will give the algorithm to reach the best solution (i.e. final ground state of the Hamiltonian, i.e. state with the minimum energy of the system). To well understand the full process in action, we will define one by one all the building block necessary to express a SAT Hamiltonian. In this paper, in section 2 we will describe the adiabatic approach from the initial principles of Schrodinger equation to generic ideas behind the AQO algorithm. Section 3 will be dedicated to SAT Model and specific experiments on simple instances. Finally, we will focus on conclusion and perspectives.

## 2. Adiabatic approaches

### 2.1. Principles

To find a solution (of a combinatorial problem specifically here), quantum algorithms require to solve the Schrödinger equation

$$\frac{\partial}{\partial t}|\psi(x,t)\rangle = -\frac{i}{\hbar}.H(t).|\psi(x,t)\rangle$$

where the energy is defined by $H(t)$ as a Hamiltonian (ie: a matrix operator).

If $H$ is time independent then the solution of this differential equation is $|\psi_t\rangle = e^{-\frac{i}{\hbar}.t.H}.|\psi_0\rangle$ where $|\psi_0\rangle$ is the initial state. In general time dependent situations, the solution is $|\psi_T\rangle = e^{-\frac{i}{\hbar}\int_o^T H(u).du}.|\psi_0\rangle$ where $T$ is the time at which the solution has to be computed.

The time interval $[0; T]$ can be divided into $n$ intervals $\Delta t = \frac{T}{n}$ in such way that $|\psi_t\rangle$ can be assumed as constant (Fig 1).

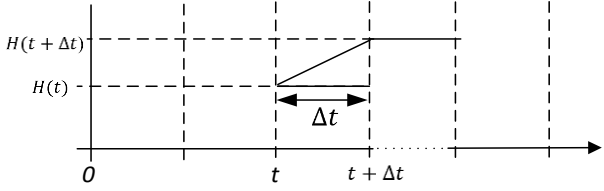

Fig. 1. $|\psi_t\rangle$ can be assumed as constant

An approximation of $\int_o^T H(u).du$ is a sum of intervals where the function can be assumed as constant:
$$\int_0^T H(u).du \simeq \sum_{k=0}^{n-1} \int_{k.\Delta t}^{(k+1).\Delta t} H(u).du$$
And by consequence:
$$|\psi_T\rangle = e^{-\frac{i}{\hbar}\int_o^T H(u).du}.|\psi_0\rangle \simeq e^{-\frac{i}{\hbar}\sum_{k=0}^{n-1}\int_{k.\Delta t}^{(k+1).\Delta t} H(u).du}.|\psi_0\rangle$$

For Hermitian operators it has been proven by (Suzuki, 1976) that
$$\lim_{n\to\infty}\left(e^{i.A.t/n}.e^{i.B.t/n}\right)^n = e^{i.(A+B).t}$$
Note that a more concise and understandable demonstration has been introduced lately by (Barthel et Zhang, 2019).

The previous remark shows that we get a good approximate expression for $e^{i.(A+B).t}$ considering:
$$e^{i.(A+B).t} \approx e^{i.(A).t}.e^{i.(B).t}$$
Now we can define:
$$|\psi_T\rangle \simeq e^{-\frac{i}{\hbar}\int_{(n-1).\Delta t}^{n\Delta t} H(u).du} \dots e^{-\frac{i}{\hbar}\int_0^{\Delta t} H(u).du}|\psi_0\rangle$$
with $$|\psi_t\rangle = e^{-\frac{i}{\hbar}\int_t^{t+\Delta t} H(u).du}.|\psi_0\rangle$$
If $H$ is time independent on $[t; t + \Delta t]$ then
$$\int_t^{t+\Delta t} H(u).du = H(t).\Delta t + O(\Delta t^2)$$
And by consequence
$$|\psi_{t+\Delta t}\rangle = e^{-\frac{i}{\hbar}[H(t).\Delta t + O(\Delta t^2)]}.|\psi_t\rangle$$
To conclude $|\psi_{t+\Delta t}\rangle = e^{-\frac{i}{\hbar}\Delta t.H(t)}.e^{-\frac{i}{\hbar}O(\Delta t^2)}|\psi_t\rangle$
and if $\Delta t$ is very small, we have: $|\psi_{t+\Delta t}\rangle \simeq e^{-\frac{i}{\hbar}\Delta t.H(t)}|\psi_t\rangle$
This Trotterization process give us insight on a way to compute the solution. It is derived by multiple smaller pieces of $|\psi_T\rangle$

According to the (Farhi, 2002) the $H(t)$ Hamiltonian can defines as a linear combination of $H_D$ to $H_P$ where $H_D$ is referred to as the "Hamiltonian Driver" and $H_P$ models the combinatorial problem.

At the first step, the system must be tuned in the ground state of one Hamiltonian $H_D$ commonly referred to the "Hamiltonian Driver" that must be defined in such a way that it does not commute with the Hamiltonian modelling the combinatorial problem. The adiabatic quantum optimization requires a Hamiltonian which is slowly tuned from $H_D$ to $H_P$ using an interpolation based on one parameter $s(t)$. Let us assume this parameter is smoothly decreasing from 1 to 0, we will define $H(t)$ as $s(t).H_D + [1 - s(t)].H_P$ (see Fig. 2).

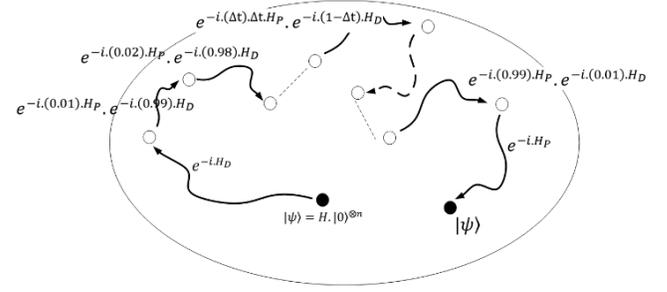

Fig. 2. Typical curve of on adiabatic process

The Ising model (Kadowaki and Nishimori, 1998) is composed of longitudinal $H_P$ and transverse field $H_D$ where $s(t)$ is the amplitude that meets the natural dynamics modeled by the Schrödinger equation. Let us note that the temperature state commonly used in Simulated Annealing (SA) and the amplitude $s(t)$ of the Adiabatic Approach play a central and similar role. A high temperature in the Simulated Annealing refers to quite similar probabilities to transition from one state to a new one without any consideration of the function landscape that leads to a mixture of states into a uniform wave function. For the adiabatic optimization, large values of $s(t)$ (close to 1) define a linear combination of states with equal amplitude. However, the quantum adiabatic evolution is quite different from cooling principles since quantum evolution enforces the system to remain in the ground state contrary to SA that generates stochastic distribution proportional to the function landscape barriers.

The solution of an optimization problem must be encoded in the ground state of a problem Hamiltonian $H_p$ that encodes the problem into the Ising model to meet the physical interpretation. The adiabatic theorem (Kato, 1950) ensures that the system stays in the ground state if the $s(t)$ time dependent evolution is achieved slowly enough to ensure that $|\psi_t\rangle$ is a state close to the ground state.

## 2.2 Algorithm
### 2.2.1 Preliminaries remarks
To derive an algorithm, considering the previous remarks, for each interval $[t; t + \Delta t]$:
$$|\psi_{t+\Delta t}\rangle = e^{-\frac{i}{\hbar}\Delta t.H(t)}|\psi_t\rangle$$
$$|\psi_0\rangle = |+\rangle^{\otimes n}$$
And at step $k$:
$$|\psi_{k+1}\rangle = e^{-\frac{i}{\hbar}\Delta t.H(k.\Delta t)}|\psi_k\rangle$$
$$|\psi_0\rangle = |+\rangle^{\otimes n}$$
For convenience a new system of units (that could present advantages) is introduced assuming that $\frac{\Delta t}{\hbar} = 1$ and leading to more compact expressions:
$$|\psi_{k+1}\rangle = e^{-i.H(k.\Delta t)}|\psi_k\rangle$$
$$|\psi_k\rangle = e^{-i.H((k-1).\Delta t)}|\psi_{k-1}\rangle$$

$|\psi_0\rangle = |+\rangle^{\otimes n}$

that also can be rewritten as:

$$|\psi_{t+\Delta t}\rangle = e^{-i.H(t)}|\psi_t\rangle$$
$$|\psi_t\rangle = e^{-i.H(t-\Delta t)}|\psi_{t-1}\rangle$$
$$|\psi_0\rangle = |+\rangle^{\otimes n}$$

The expression $|\psi_{t+\Delta t}\rangle = e^{-\frac{i}{\hbar}\Delta t.H(t)}|\psi_t\rangle$ can be rewritten considering that $e^x = 1 + x + O(x^2)$:

$$|\psi_{t+\Delta t}\rangle = e^{-i.H(t)}|\psi_t\rangle$$
$$|\psi_{t+\Delta t}\rangle = [1 - i.H(t) + O([-i.H(t)]^2)]|\psi_t\rangle$$
$$|\psi_{t+\Delta t}\rangle = [1 - i.H(t)].|\psi_t\rangle + O(\Delta t^2)$$
$$|\psi_{t+\Delta t}\rangle = |\psi_t\rangle - i.H(t).|\psi_t\rangle + O(\Delta t^2)$$

Let us assume that $|\psi_t\rangle$ is the eigenvector of $H(t)$ with the eigenvalue $\lambda_t$:

$$|\psi_{t+\Delta t}\rangle = |\psi_t\rangle - i.\lambda_t.|\psi_t\rangle + O(\Delta t^2)$$
$$|\psi_{t+\Delta t}\rangle = (1 - i.\lambda_t).|\psi_t\rangle + O(\Delta t^2)$$

Here we introduce a way to compute each piece of $|\psi_T\rangle$ by recurrence from the previous piece. Knowing an initial state and this relation we can now derive an algorithm.

This last expression proves that $|\psi_{t+\Delta t}\rangle$ is a combination of $|\psi_t\rangle$ minus $i.\lambda_t.|\psi_t\rangle$ which is orthogonal to $|\psi_t\rangle$ since $i$ is a rotation of 90°. $O(\Delta t^2)$ is the difference between $|\psi_{t+\Delta t}\rangle$ and $|\psi_t\rangle$ as stressed in the figure 3.

### 2.2.2 Key-points

One proposal to compute this adiabatic process can be achieved (Farhi et al., 2000) (Farhi et al., 2002) and (Farhi et al., 2014) by considering only very specific Hamiltonians

$$H(t) = s(t).H_D + [1 - s(t)].H_P$$

where $s(t) = t$ and where $t$ refers to the intervals of figure 2.

$$H(t) = t.H_D + (1 - t).H_P$$

with
$$H(t + \Delta t) = (t + \Delta t).H_D + (1 - t - \Delta t).H_P$$
$$H(t + \Delta t) = H(t) + \Delta t.(H_D - H_P)$$

Thus
$$H(t + \Delta t).|\psi_{t+\Delta t}\rangle = [H(t) + \Delta t.(H_D - H_P)].[(1 - i.\lambda_t).|\psi_t\rangle + O(\Delta t^2)]$$
$$H(t + \Delta t).|\psi_{t+\Delta t}\rangle = H(t).(1 - i.\lambda_t).|\psi_t\rangle + \Delta t.(H_D - H_P).(1 - i.\lambda_t).|\psi_t\rangle + O(\Delta t^2)$$

Assuming $|\psi_t\rangle$ is an eigenvector of $H_t$ with an eigenvalue $\lambda_t$:
$$H(t + \Delta t).|\psi_{t+\Delta t}\rangle = \lambda_t.(1 - i.\lambda_t).|\psi_t\rangle + \Delta t.(H_D - H_P).(1 - i.\lambda_t).|\psi_t\rangle + O(\Delta t^2)$$
$$H(t + \Delta t).|\psi_{t+\Delta t}\rangle = \lambda_t.(1 - i.\lambda_t).|\psi_t\rangle + \Delta t.(H_D - H_P)|\psi_t\rangle - \Delta t.(H_D - H_P).i.\lambda_t|\psi_t\rangle + O(\Delta t^2)$$
$$H(t + \Delta t).|\psi_{t+\Delta t}\rangle = \lambda_t.(1 - i.\lambda_t).|\psi_t\rangle + \Delta t.(H_D - H_P).|\psi_t\rangle + O(\Delta t^2)$$

Since $\Delta t.(H_D - H_P).i.\lambda_t|\psi_t\rangle$ is in the order to $\Delta t$:
$$H(t + \Delta t).|\psi_{t+\Delta t}\rangle = \lambda_t.|\psi_{t+\Delta t}\rangle + O(\Delta t)$$

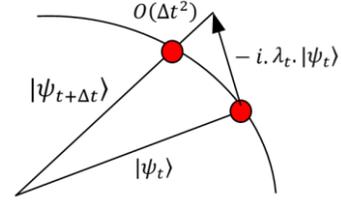

**Fig. 3. Typical curve of on adiabatic process**

$|\psi_{t+\Delta t}\rangle$ is an eigenvector of $H(t + \Delta t)$ with the eigenvalue $\lambda_t$ if $O(\Delta t)$ is small enough meaning that $|\psi_{t+\Delta t}\rangle$ and $|\psi_t\rangle$ lead to the same energy level (related to minimal spectral gap).

All the previous key points (initial state, recurrence form, small value of $\Delta t$) can help us to derive a computation of $|\psi\rangle$ as described in Algorithm 1.

### 2.2.3. Initialization and iterations

The initial state is obtained by application of the Hadamard gate $H$ to $|0\rangle^{\otimes n}$ and the initial state is $|\psi_0\rangle = H.|0\rangle^{\otimes n}$

One knows that
$$X|+\rangle = |+\rangle$$
$$X|-\rangle = -|-\rangle$$

meaning that the eigenvalue 1 refers to the eigenvector $|+\rangle$ and that the eigenvalue $-1$ refers to the eigenvector $|-\rangle$ considering $|+\rangle = H|0\rangle$ and $|-\rangle = H|1\rangle$). Because the initial Hamiltonian is only directed on $H_D$ the minimal eigenvalue $-1$ must be assigned to $|+\rangle$. The usual local transverse-field Hamiltonian is

$$H_D = \sum_{i=1}^{n} -X_i$$

with a strength that varies with respect to time and where $X_i$ is the $X$ operator acting on the $i^{th}$ qubit where the ground state $H_0$ is the superposition of all possible states in the eigenbasis of $H_P$.

The eigenvalues of $H_D$ are 1 and $-1$ and the state $|+\rangle^{\otimes n}$ refers to the eigenvalue $-1$ defining the fundamental ground state $H_D$.

One iteration $j$ consists in operator $e^{-i.(1-s_j).H_P}.e^{-i.s_j.H_D}$ where $s_j$ is a parameter smoothly varying from 1 ($s_0 = 1$) to 0. Subsequently, on first iteration the algorithm generates from initial state $|\psi_0\rangle = H.|0\rangle^{\otimes n}$ a new state $|\psi_1\rangle$ by application of operator $e^{-i.\Delta t.H_P}.e^{-i.(1-\Delta t).H_D}$ to $|\psi_0\rangle$.

From $t = 0$ the new state is:
$$|\psi_1\rangle = e^{-i.0.\Delta t.H_P}.e^{-i.(1-0).H_D}.|\psi_0\rangle$$
$$|\psi_1\rangle = e^{-i.H_D}.|\psi_0\rangle$$

Similarly, at the second iteration, the algorithm generates from $|\psi_1\rangle$ a new state $|\psi_2\rangle$ by application of the operator $e^{-i.(1.\Delta t).H_P}.e^{-i.(1-.1.\Delta t).H_D}$ to $|\psi_1\rangle$ leading to
$$|\psi_2\rangle = e^{-i.\Delta t.H_P}.e^{-i.(1-\Delta t).H_D}.|\psi_1\rangle$$

And so one until we reach the last iteration where
$$|\psi_T\rangle = e^{-i.1.H_P}.e^{-i.0.H_D}.|\psi_{T-1}\rangle$$

**Adiabatic Algorithm**

    Input Parameters:
        $H_D$ : Hamiltonian mixer
        $H_P$ : Hamiltonian
        $\Delta t$ : A small float value (for example $\Delta t = 1/T$)
        $T$ : number of intervals (iterations) in Schrödinger equation simulation
    Output parameters:
        $S^*$ : best solution found in sampling of $|\psi\rangle$
    Local parameters:
        $j$ : iterations
    Begin

1. $|\psi(s)\rangle = H.|0\rangle^{\otimes n}$
2. For j from 0 to T do
3. $s(j) = j.\Delta t$
4. $|\psi(s)\rangle = e^{-i.(1-s(j)).H_D}.|\psi(s)\rangle$
5. $|\psi(s)\rangle = e^{-i.s(j).H_P}.|\psi(s)\rangle$
6. End

Considering $|\psi(s)\rangle$ compute state $S^*$ which has the largest probability in the wave function of $|\psi\rangle$.

End

**Algorithm 1: Adiabatic Quantum Algorithm**

## 3. SAT model and resolution

SAT problem was the first decision problem to be demonstrated as NP-complete. The 3-SAT problem is a subproblem of deciding whether a given propositional Boolean formula expressed as a conjunction of clauses consisting of at most three literals is satisfiable. The Max-SAT problem for a formula $\varphi$ is the problem of finding an assignment of values to variables that minimizes the number of unsatisfied clauses in $\varphi$.

We propose to analyze the process of the adiabatic algorithm on SAT instances to challenge conviction in the advantage of such quantum algorithms. Note that numerical experiments will be focused only on very small size to favor explanation of the model and analysis of the results. However, the current experiments have been achieved on the simulator provided by IBM that allow useful abstraction for dealing with quantum computers and achieving state-of-the-art results.

The Hamiltonian of a system fully describes the dynamics. As the Hamiltonian is a Hermitian operator it has a spectral decomposition: $H = \sum_i \lambda_i |e_i\rangle\langle e_i|$ which means: $H$ is a diagonalized operator (where $\lambda_i$ are the eigenvalues and $|e_i\rangle$ are the basis vectors, representative of the observable measure).

### 3.1. Projections and Operators

A SAT formula is modeled with a Hamiltonian where we use for each clause a Hamiltonian connected with controlled-unitary operator. It computes function value in a qubit register. Boolean functions are modeled by a linear combination of Pauli Z operators. Every operator can be described in a tensor representation or a combination of projectors, or in a geometric view.

We have the operator $Z_j$ ($j = 1, n$) by definition:

$$Z_j = \overbrace{Id \otimes Id \dots \otimes Id}^{j-1} \otimes Z \otimes \overbrace{Id \otimes Id \dots \otimes Id}^{n-j}$$

and let us consider a state:

$$|u\rangle = \overbrace{|u_1\rangle \otimes |u_2\rangle \otimes \dots \otimes |u_{j-1}\rangle}^{j-1} \otimes |0\rangle \otimes \overbrace{|u_{j+1}\rangle \otimes \dots \otimes |u_n\rangle}^{n-j}$$

Hence

$$Z_j.|u\rangle = \overbrace{|u_1\rangle \otimes |u_2\rangle \otimes \dots \otimes |u_{j-1}\rangle}^{j-1} \otimes Z.|0\rangle \otimes \overbrace{|u_{j+1}\rangle \otimes \dots \otimes |u_n\rangle}^{n-j}$$

$$Z_j.|u\rangle = \overbrace{|u_1\rangle \otimes |u\rangle \otimes \dots \otimes |u_{j-1}\rangle}^{j-1} \otimes |0\rangle \otimes \overbrace{|u_{j+1}\rangle \otimes \dots \otimes |u_n\rangle}^{n-j}$$

$$Z_j.|u\rangle = |u\rangle$$

Similarly with

$$|u\rangle = \overbrace{|u_1\rangle \otimes |u_2\rangle \otimes \dots \otimes |u_{j-1}\rangle}^{j-1} \otimes |1\rangle \otimes \overbrace{|u_{j+1}\rangle \otimes \dots \otimes |u_n\rangle}^{n-j}$$

We have:

$$Z_j.|u\rangle = \overbrace{|u_1\rangle \otimes |u_2\rangle \otimes \dots \otimes |u_{j-1}\rangle}^{j-1} \otimes Z.|1\rangle \otimes \overbrace{|u_{j+1}\rangle \otimes \dots \otimes |u_n\rangle}^{n-j}$$

$$Z_j.|u\rangle = \overbrace{|u_1\rangle \otimes |u_2\rangle \otimes \dots \otimes |u_{j-1}\rangle}^{j-1} \otimes (-|1\rangle) \otimes \overbrace{|u_{j+1}\rangle \otimes \dots \otimes |u_n\rangle}^{n-j}$$

That is $Z_j.|u\rangle = -|u\rangle$ illustrating that $Z_j$ swap the qubit in the $j^{th}$ position if the qubit value is $|1\rangle$.

The operators $Z_j$ can be described using projectors $|e_i\rangle\langle e_i|$ that are the projection on the $i^{th}$ basis vector.

In the specific situation where $n = 1$

$$Z = Z_1 = 2.|e_1\rangle.\langle e_1| - Id$$
In the specific situation where $n = 2$
$$Z_1 = 2.|e_1\rangle.\langle e_1| + 2.|e_2\rangle.\langle e_2| - Id^{\otimes 2}$$
$$Z_2 = 2.|e_1\rangle.\langle e_1| + 2.|e_3\rangle.\langle e_3| - Id^{\otimes 2}$$
In the specific situation where $n = 3$
$$Z_1 = 2.|e_1\rangle.\langle e_1| + 2.|e_2\rangle.\langle e_2| + 2.|e_3\rangle.\langle e_3| + 2.|e_4\rangle.\langle e_4| - Id^{\otimes 3}$$
$$Z_2 = 2.|e_1\rangle.\langle e_1| + 2.|e_2\rangle.\langle e_2| + 2.|e_5\rangle.\langle e_5| + 2.|e_6\rangle.\langle e_6| - Id^{\otimes 3}$$
$$Z_3 = 2.|e_1\rangle.\langle e_1| + 2.|e_3\rangle.\langle e_3| + 2.|e_5\rangle.\langle e_5| + 2.|e_7\rangle.\langle e_7| - Id^{\otimes 3}$$

When $n = 1$, $Z_1$ is the Pauli operator applied to $|x\rangle$ is illustrated on figure 4.

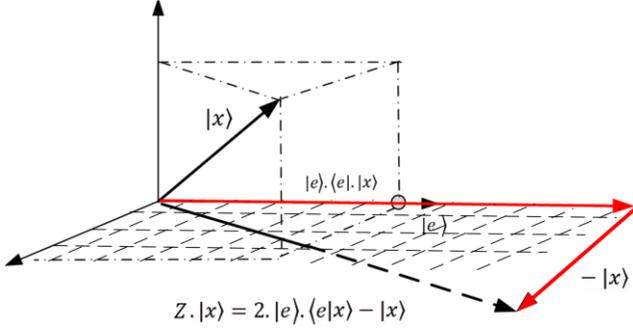

**Fig. 4. Visualization of $Z.|x\rangle$ $(n = 1)$**

Let us denote $M = diag(m_1, m_2, \dots, m_n)$ the diagonal matrix $M$ where diagonal terms are $m_1, m_2, \dots, m_n$.
But, if $Id = diag(1,1)$ then
$$Id \otimes M = diag(m_1, m_2, \dots, m_n, m_1, m_2, \dots, m_n)$$
and
$$M \otimes Id = diag(m_1, m_1, m_2, m_2, \dots, m_n, m_n)$$

Consequently, for example with $n = 2$, for $Z = diag(1, -1)$:
$$Z_1 = Z \otimes Id = diag(1, 1, -1, -1)$$
and
$$Z_2 = Id \otimes Z = diag(1, -1, 1, -1)$$

With $n = 3$,
$$Z_1 = (Z \otimes Id) \otimes Id = diag(1, 1, -1, -1) \otimes Id$$
$$Z_1 = (Z \otimes Id) \otimes Id = diag(1, 1, 1, 1, -1, -1, -1, -1)$$

$$Z_2 = Id \otimes (Z \otimes Id) = Id \otimes diag(1, 1, -1, -1)$$
$$Z_2 = Id \otimes (Z \otimes Id) = diag(1, 1, -1, -1, 1, 1, -1, -1)$$
and
$$Z_3 = Id \otimes (Id \otimes Z) = Id \otimes diag(1, -1, 1, -1)$$
$$Z_3 = Id \otimes (Id \otimes Z) = diag(1, -1, 1, -1, 1, -1, 1, -1)$$

In the qubit base $|e_k\rangle_{1 \le k \le q}$, for any $|x\rangle = \sum_{k=1}^{q} x_k.|e_k\rangle$ we have $|e_j\rangle.\langle e_j|.|x\rangle = |e_j\rangle.\langle e_j|x\rangle = |e_j\rangle.x_j$ By consequence, $Z_j$ is an operator that can be defined considering the next formula where $q = 2^n$:

$$Z_j = 2 \sum_{u=1}^{2^{j-1}} \sum_{v=1}^{\left(\frac{2^n}{2^j}\right)} \left|e_{u \times \left(\frac{2^n}{2^{j-1}}\right)+v-1}\right\rangle \left\langle e_{u \times \left(\frac{2^n}{2^{j-1}}\right)+v-1}\right| - Id^{\otimes n}$$

$$Z_j = 2 \sum_{u=1}^{2^{j-1}} \sum_{v=1}^{2^{n-j}} |e_{u \times 2^{n-j+1}+v-1}\rangle \langle e_{u \times 2^{n-j+1}+v-1}| - Id^{\otimes n}$$

$$Z_j = 2 \sum_{k=1}^{2^{j-1}} \left( \sum_{v=(k-1) \times 2^{n-j+1}}^{(k-1) \times 2^{n-j+1}+2^{n-j}-1} |e_v\rangle \langle e_v| \right) - Id^{\otimes n}$$

For example, when $n = 4$ ($q = 2^n = 2^4 = 16$) we have:

$$Z_1 = 2 \sum_{k=1}^{1} \left( \sum_{j=0}^{7} |e_j\rangle \langle e_j| \right) - Id^{\otimes n}$$

and by consequence
$$Diag(Z_1) = (1,1,1,1,1,1,1,1,-1,-1,-1,-1,-1,-1,-1,-1)$$

$$Z_2 = 2 \sum_{k=1}^{2} \left( \sum_{v=(k-1) \times (2^3)}^{(k-1) \times (2^3)+(2^2)-1} |e_v\rangle \langle e_v| \right) - Id^{\otimes n}$$

$$Z_2 = 2 \left( \left( \sum_{v=0}^{3} |e_j\rangle \langle e_j| \right) + \left( \sum_{v=8}^{8+3=11} |e_v\rangle \langle e_v| \right) \right) - Id^{\otimes 4}$$

and by consequence
$$Diag(Z_2) = (1,1,1,1,-1,-1,-1,-1,1,1,1,1,-1,-1,-1,-1)$$

$$Z_3 = 2 \sum_{k=1}^{(2^2)} \left( \sum_{v=(k-1) \times (2^2)}^{(k-1) \times (2^2)+(2^1)-1} |e_v\rangle \langle e_v| \right) - Id^{\otimes n}$$

$$Z_3 = 2 \left( \sum_{v=0}^{1} |e_v\rangle \langle e_v| + \sum_{v=4}^{5} |e_v\rangle \langle e_v| + \sum_{j=8}^{9} |e_v\rangle \langle e_v| + \sum_{j=12}^{13} |e_v\rangle \langle e_v| \right) - Id^{\otimes 4}$$

and by consequence
$$Diag(Z_3) = (1,1,-1,-1,1,1,-1,-1,1,1,-1,-1,1,1,-1,-1)$$

$$Z_4 = 2 \sum_{k=1}^{(2^3)} \left( \sum_{v=(k-1) \times (2^1)}^{(k-1) \times (2^1)+(2^0)-1} |e_v\rangle \langle e_v| \right) - Id^{\otimes n}$$

$$Z_4 = 2.\left( \sum_{j=0}^{0} |e_j\rangle \langle e_j| + \sum_{j=2}^{2} |e_j\rangle \langle e_j| + \dots + \sum_{j=12}^{12} |e_j\rangle \langle e_j| + \sum_{j=14}^{14} |e_j\rangle \langle e_j| \right) - Id^{\otimes 4}$$

and by consequence
$$Diag(Z_4) = (1,-1,1,-1,1,-1,1,-1,1,-1,1,-1,1,-1,1,-1)$$

To conclude we can confirm that $Z_j$ transforms $|x\rangle$ into a new symmetric state:

$$Z_j = 2 \sum_{k=1}^{2^{j-1}} \left( \sum_{v=(k-1)\times 2^{n-j+1}}^{(k-1)\times 2^{n-j+1}+2^{n-j}-1} |e_v\rangle\langle e_v|x\rangle \right) - |x\rangle$$

achieving an orthogonal projection of $|x\rangle$ on the direction of $|e_v\rangle$ where $v = (k-1) \times \left(\frac{2^n}{2^{j-1}}\right) .. (k-1) \times \left(\frac{2^n}{2^{j-1}}\right) + \left(\frac{2^n}{2^j}\right) - 1$ and for all value of $k$ such that $k = 1 \ldots 2^{j-1}$.

This formula can be rewritten

$$Z_j = 2 \sum_{v \notin E_{Z_j}} |e_v\rangle\langle e_v|x\rangle - |x\rangle$$

With
$E_{Z_j}$
$= \begin{cases} j \;/\; k \notin \left[1 .. \frac{2^n}{2^{n-j+1}}\right] \text{ or} \\ v \notin i\left[(k-1) \times \left(\frac{2^n}{2^{j-1}}\right), (k-1) \times \left(\frac{2^n}{2^{j-1}}\right) + \left(\frac{2^n}{2^j}\right) - 1\right] \end{cases}$

$Z_j. |x\rangle$ is referred to as

$$P_{\perp Z_j}(|x\rangle) = 2 \sum_{k=1}^{2^{j-1}} \left( \sum_{v=(k-1)\times 2^{n-j+1}}^{(k-1)\times 2^{n-j+1}+2^{n-j}-1} |e_v\rangle\langle e_v|x\rangle \right)$$

and by consequence

$$|x\rangle - P_{\perp Z_j}(|x\rangle) = |x\rangle - 2 \sum_{k=1}^{2^{j-1}} \left( \sum_{v=(k-1)\times 2^{n-j+1}}^{(k-1)\times 2^{n-j+1}+2^{n-j}-1} |e_v\rangle\langle e_v|x\rangle \right)$$

is the difference between $|x\rangle$ and the projection of $|x\rangle$ on all the orthogonal planes.

To conclude
$$Z_j|e_k\rangle = \begin{cases} -1 \text{ if } k \in E_{Z_j} \\ +1 \text{ if } k \notin E_{Z_j} \end{cases}$$

with
$E_{Z_j} = \left\{ j \;/\; k \notin \left[1 .. \frac{2^n}{2^{n-j+1}}\right] \text{ or } v \notin i\left[(k-1) \times \left(\frac{2^n}{2^{j-1}}\right), (k-1) \times \left(\frac{2^n}{2^{j-1}}\right) + \left(\frac{2^n}{2^j}\right) - 1\right] \right\}$. In other words, $E_{Z_j}$ is the set of positions where the eigenvalue is $-1$. For example we have $E_{Z_3} = \{2; 4; 6; 8\}$. We can state that $Z_j$ operator over any basis state will revert every $e_k / k \in E_{Z_j}$.

### 3.2. Specific Hamiltonian with 2 qubits.
The Hamiltonian dependent problem $H_P$ can be decomposed with $Z_i$ gates leading to, for example, $H_P = \omega.Id + \alpha.Z_i + \beta.Z_j + \gamma.Z_i.Z_j$.

Considering $Z_i$ gate only, we have
$$\frac{d|\psi_t\rangle}{dt} = -i.H.|\psi_t\rangle$$
that can be rewritten:

$$\frac{d|\psi_t\rangle}{dt} = [-i.(\omega.Id + \alpha.Z_i + \beta.Z_j + \gamma.Z_i.Z_j)].|\psi_t\rangle$$
$$\frac{d|\psi_t\rangle}{dt} = (-i\omega.Id - i.\alpha.Z_i - i.\beta.Z_j - i.\gamma.Z_i.Z_j).|\psi_t\rangle$$

The solution of
$$\frac{d|\psi_t\rangle}{dt} = (-i\omega.Id - i.\alpha.Z_i - i.\beta.Z_j - i.\gamma.Z_i.Z_j).|\psi_t\rangle$$
Is:
$$|\psi_t\rangle = e^{-i.\omega.t.Id - i.t.\alpha.Z_i - i.t.\beta.Z_j - i.t.\gamma.Z_i.Z_j}.|\psi_0\rangle$$
$$|\psi_t\rangle = e^{-i.t.\omega.Id}.e^{-i.t.\alpha.Z_i}.e^{-i.t.\beta.Z_j}.e^{-i.t.\gamma.Z_1.Z_j}.|\psi_0\rangle$$
$$|\psi_t\rangle = e^{-i.t.\omega}.R_Z^i(t.\alpha).R_Z^j(t.\beta).e^{-i.t.\gamma.Z_i.Z_j}.|\psi_0\rangle$$

Because $e^{-i.t.\omega}$ does not impact $|\psi_t\rangle$, then $e^{-i.t.\omega}$ can be removed to obtain a more compact expression:
$$|\psi_t\rangle = R_Z^i(t.\alpha).R_Z^j(t.\beta).e^{-i.t.\gamma.Z_i.Z_j}.|\psi_0\rangle$$

If $i = 1$ and $j = 2$ then we can note $R_Z^1(t.\alpha).R_Z^2(t.\beta)$ that leads to the formula:
$$R_Z^1(t.\alpha).R_Z^2(t.\beta) = [R_Z(t.\alpha) \otimes id].[id \otimes R_Z(t.\beta)]$$
$$= R_Z(t.\alpha) \otimes R_Z(t.\beta).$$

### 3.3. Matrix expression and circuits
For any operator $U$ such $U^2 = Id$ we have:
$$e^{i.\alpha.U} = \sum_{k=0}^{k=\infty} \frac{(i.\alpha.U)^k}{k!}$$
$$e^{i.\alpha.U} = \cos(\alpha).Id + i.\sin(\alpha).U$$
and by consequence, over 2 qubits when Application of $Z = \begin{pmatrix} 1 & 0 \\ 0 & -1 \end{pmatrix}$ on the first qubit of $|\phi\rangle \otimes |\psi\rangle$ is defined by the operator
$$Z_1 = Z \otimes Id$$
and the application of $Z$ on the qubit 2 is $Z_2 = Id \otimes Z$, we have :
$$e^{-i.t.\alpha.Z_1} = \cos(t.\alpha).Id - i.\sin(t.\alpha).Z_1$$

$$e^{-i.t.\alpha.Z_1} = \cos(t.\alpha).\begin{pmatrix} 1 & 0 & 0 & 0 \\ 0 & 1 & 0 & 0 \\ 0 & 0 & 1 & 0 \\ 0 & 0 & 0 & 1 \end{pmatrix}$$
$$-i.\sin(t.\alpha).\begin{pmatrix} 1 & 0 & 0 & 0 \\ 0 & 1 & 0 & 0 \\ 0 & 0 & -1 & 0 \\ 0 & 0 & 0 & -1 \end{pmatrix}$$

and
$$e^{-i.t.\alpha.Z_2} = \cos(t.\alpha).\begin{pmatrix} 1 & 0 & 0 & 0 \\ 0 & 1 & 0 & 0 \\ 0 & 0 & 1 & 0 \\ 0 & 0 & 0 & 1 \end{pmatrix}$$
$$-i.\sin(t.\alpha).\begin{pmatrix} 1 & 0 & 0 & 0 \\ 0 & -1 & 0 & 0 \\ 0 & 0 & 1 & 0 \\ 0 & 0 & 0 & -1 \end{pmatrix}$$

Because $(Z_1.Z_2)^2 = Id^{\otimes 2}$
$$e^{-i.t.\gamma.Z_1.Z_2} = \cos(t.\gamma).Id - i.\sin(t.\gamma).Z_1.Z_2$$

$$e^{-i.t.\gamma.Z_1.Z_2} = \cos(t.\gamma) \cdot \begin{pmatrix} 1 & 0 & 0 & 0 \\ 0 & 1 & 0 & 0 \\ 0 & 0 & 1 & 0 \\ 0 & 0 & 0 & 1 \end{pmatrix}$$

$$- i.\sin(t.\gamma) \cdot \begin{pmatrix} 1 & 0 & 0 & 0 \\ 0 & -1 & 0 & 0 \\ 0 & 0 & -1 & 0 \\ 0 & 0 & 0 & 1 \end{pmatrix}$$

Since $\cos(t.\gamma) - i.\sin(t.\gamma) = e^{-i.t.\gamma}$ and $\cos(t.\gamma) + i.\sin(t.\gamma) = e^{i.t.\gamma}$ we have:

$$e^{-i.t.\gamma.Z_1.Z_2} = \begin{pmatrix} e^{-i.t.\gamma} & 0 & 0 & 0 \\ 0 & e^{i.t.\gamma} & 0 & 0 \\ 0 & 0 & e^{i.t.\gamma} & 0 \\ 0 & 0 & 0 & e^{-i.t.\gamma} \end{pmatrix}$$

The circuit that performs $e^{-i.t.\gamma.Z_1.Z_2}$ is introduced in figure 5 (Hadfield, 2021)

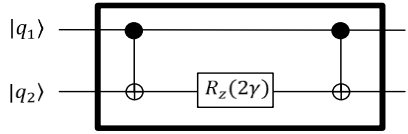

**Fig 5. Quantum circuit performing $e^{-i.t.\gamma.Z_1.Z_2}$**

The circuit $CX_{1,2}.R_Z^2(2t.\gamma).CX_{1,2}$ is the modelization of $e^{-i.t.\gamma.Z_1.Z_2}$. It can be easily verified by applying definition of $CX$ over first and second qubits and $R_Z^2$. Similarly we can see than the quantum circuit $CX_{1,2}.CX_{2,3}.R_Z^2(2t.\gamma).CX_{2,3}.CX_{1,2}$ performs $e^{-i.t.\gamma.Z_1.Z_2.Z_3}$

### 3.4. Specific Hamiltonian for Boolean expressions.

An efficient methodological way to represent basic Boolean features (literal) of SAT must permit to model both conjunctions (AND) and disjunctions (OR). One function $f = 0$ and $f = 1$ can be modeled with the Hamiltonian: $H_{x_j} = \frac{1}{2}(Id^{\otimes n} - Z_j)$

A basic application of propositional logic law, permits to derive rules for composing Hamiltonians representing clauses and to construct representations of clauses or functions including Boolean formulas. The recursive construction rules have been first defined by (Hadfield, 2021) but it is possible to have an intuitive way to understand the model based on $\frac{1}{2}(Id^{\otimes n} - Z_j)$ by a cautious analysis of the operator. Note that in the very specific situation where $n = 1$, we have: $\frac{1}{2}(Id - Z)$

Let us consider $f$ such that, for $k \in E_{Z_j}, f(e_k) = 1$ else $f(e_k) = 0$ with $\frac{1}{2}(Id^{\otimes n} - Z_j)$

Because

$$Z_j|e_k\rangle = \begin{cases} -1 \text{ if } k \in E_{Z_j} \\ +1 \text{ if } k \notin E_{Z_j} \end{cases}$$

we have

$$(Id^{\otimes n} - Z_j)|e_k\rangle = \begin{cases} 2 \text{ if } k \in E_{Z_j} \\ 0 \text{ if } k \notin E_{Z_j} \end{cases}$$

and

$$\left(\frac{Id^{\otimes n} - Z_j}{2}\right)|e_k\rangle = \begin{cases} 1 \text{ if } k \in E_{Z_j} \\ 0 \text{ if } k \notin E_{Z_j} \end{cases}$$

By consequence, for any $|x\rangle = \sum_k x_k.|e_k\rangle$ we have

$$\left(\frac{Id^{\otimes n} - Z_j}{2}\right)|x\rangle = \sum_{k \in E_{Z_j}} x_k.|e_k\rangle$$

For

$$\left(\frac{Id^{\otimes n} - Z_j}{2}\right)|e_k\rangle = f(e_k)|e_k\rangle$$

and

$$\left\langle x\left|\left(\frac{Id^{\otimes n} - Z_j}{2}\right)\right|x\right\rangle = \sum_{k' \in E_{Z_j}} \overline{x_{k'}} \cdot \left\langle e_{k'}\left|\sum_{k \in E_{Z_j}} x_k.|e_k\rangle\right.\right\rangle$$

$$\left\langle x\left|\left(\frac{Id^{\otimes n} - Z_j}{2}\right)\right|x\right\rangle = \sum_{k' \in E_{Z_j}} \overline{x_{k'}} \cdot \sum_{k \in E_{Z_j}} x_k \langle e_{k'}|e_k\rangle$$

$$\left\langle x\left|\left(\frac{Id^{\otimes n} - Z_j}{2}\right)\right|x\right\rangle = \sum_{k \in E_{Z_j}} \overline{x_k}.x_k \langle e_k|e_k\rangle = \sum_{k \in E_{Z_j}} |x_k|^2.$$

This formula proves that minimization of $\left\langle x\left|\left(\frac{Id^{\otimes n} - Z_j}{2}\right)\right|x\right\rangle$ is similar for the minimization of $\sum_{k \in E_{Z_j}} |x_k|^2$ and the minimal value is $\sum_{k \in E_{Z_j}} |x_k|^2 = 0$ i.e. for $x_k = 0$ if $k \in E_{Z_j}$.

**Proposition 1**
In the qubit basis $|0\rangle$ is used to model true and $|1\rangle$ to model false. By consequence, a binary variable which can have only two values (true and false) is modelized by $|0\rangle$ and $|1\rangle$. A true binary variable is efficiently modeled by $H = -Z = Diag_H(-1,1)$ which has 2 eigenvalues 1 and $-1$. Because the minimal eigenvalue is concerned with the eigenvector $|0\rangle$, one can conclude that $-Z$ is the correct modelization of a binary variable assigned to true.
Similarly a false binary variable is efficiently modeled by $H = Z = Diag_H(1,-1)$ with the eigenvalue 1 assigned to $|0\rangle$ and the eigenvalue 1 assigned to $|1\rangle$, proving that $Z$ is the correct modelization of a binary variable which value false.

The construction of Hamiltonian dedicated to Boolean function is the direct consequence of the proposition 1. The mapping to the binary variable is obtained considering $H = \frac{1}{2}Id - \frac{1}{2}Z$ which has two eigenvalues 0 and 1 that model $|0\rangle$ and $H = \frac{1}{2}Id + \frac{1}{2}Z$ which has the eigenvalue 1 and 0, that model $|1\rangle$.

The formal rules that define the correct combinations of Hamiltonian to model Boolean functions have been lately introduced in (Hadfield, 2018) and they permit to defined a set of basic Hamiltonians. For each combination of simple expression with one literal or two literals, we apply tensor product and express the associated H, giving the expression as a Diag form and the associated minimal eigenvalue(s).

Application of the rules to two binary variables can be achieved considering that:
$$Z_1 = Z \otimes Id = \text{Diag}(1,1,-1,-1)$$
$$Z_2 = Id \otimes Z = \text{Diag}(1,-1,1,-1)$$
And by consequence
$$Z_1 + Z_2 = \text{Diag}(2,0,0,-2),$$
$$-Z_1 - Z_2 = \text{Diag}(-2,0,0,2),$$
$$Z_1 - Z_2 = \text{Diag}(0,2,-2,0)$$
$$Z_1.Z_2 = Z \otimes Z = \text{Diag}(1,-1,-1,1).$$
Thus for example the operator
$$H = -Z_1 - Z_2 + Z_1.Z_2 = \text{Diag}(-1,-1,-1,3)$$
has two eigenvalues (3 and $-1$) and the ground states are $|00\rangle, |10\rangle, |01\rangle$ corresponding to $q_1.q_2 = 0$ with $q_1$ and $q_2$ being the two binary variables. All basic Boolean operations can be simulated in a similar way.

Because the Pauli $Z$ gate commutes, circuits for individual terms in the Hamiltonians can be applied in sequence to simulate their sum.

### 3.5. SAT example
Let us consider the following SAT problem $\varphi$ composed of 4 clauses and 3 variables:
$$(x_1 \lor x_2 \lor \overline{x_3})$$
$$(\overline{x_1} \lor \overline{x_2} \lor \overline{x_3})$$
$$(\overline{x_1} \lor x_2 \lor x_3)$$
$$(\overline{x_1} \lor x_2 \lor \overline{x_3})$$
Let us consider the following assignments that define solutions of the problem: (0,1,0), (0,0,0), (1,1,0), (0,1,1). In the ket basis $|0\rangle$ is used to model true (1 in the classical binary representation) and $|1\rangle$ is used to model false (0 in the classical binary representation). By consequence, the 4 solutions of the 3-SAT problem refer to $(|1\rangle |0\rangle |1\rangle), (|1\rangle |1\rangle |1\rangle), (|0\rangle |0\rangle |1\rangle), (|1\rangle |0\rangle |0\rangle)$.

Modeling of the SAT derives from the previous rules and permits mapping from $\varphi$ to Pauli $Z$ gate allows us to obtain the Hamiltonians corresponding to the $\varphi$ clauses. A full detailed description in available in chapter 6 page 322 (Bourreau et al., 2022).
$$C_1 = \frac{1}{8}Id + \frac{1}{8}(-Z_1 - Z_2 + Z_3 + Z_1.Z_2 - Z_1.Z_3 - Z_2.Z_3 + Z_1.Z_2.Z_3)$$
$$C_2 = \frac{1}{8}Id + \frac{1}{8}(Z_1 + Z_2 + Z_3 + Z_1.Z_2 + Z_1.Z_3 + Z_2.Z_3 + Z_1.Z_2.Z_3)$$
$$C_3 = \frac{1}{8}Id + \frac{1}{8}(Z_1 - Z_2 - Z_3 - Z_1.Z_2 - Z_1.Z_3 + Z_2.Z_3 + Z_1.Z_2.Z_3)$$
$$C_4 = \frac{1}{8}Id + \frac{1}{8}(Z_1 - Z_2 + Z_3 - Z_1.Z_2 + Z_1.Z_3 - Z_2.Z_3 - Z_1.Z_2.Z_3)$$
Because clause 1 and clause 2 must hold simultaneously ($H_{12} = (C_1 + C_2)$) the Hamiltonian is:

$$H_{12} = \frac{1}{4}Id + \frac{1}{4}Z_3 + \frac{1}{4}Z_1.Z_2 + \frac{1}{4}Z_1.Z_2.Z_3$$
The following Hamiltonian ensures that clause 3 is true with clause 1 and 2 ($H_{12}$).
$$H_{123} = \frac{3}{8}Id + \frac{1}{8}Z_1 - \frac{1}{8}Z_2 + \frac{1}{8}Z_3 + \frac{1}{8}Z_1.Z_2 - \frac{1}{8}Z_1.Z_3 + \frac{1}{8}Z_2.Z_3 + \frac{3}{8}Z_1.Z_2.Z_3$$
Finally, the Hamiltonian that ensures that the four clauses simultaneously hold is:
$$H_P = \frac{1}{2}Id + \frac{1}{4}Z_1 - \frac{1}{4}Z_2 + \frac{1}{4}Z_3 + \frac{1}{4}Z_1.Z_2.Z_3$$

A large enough value of $T$ must be used in the adiabatic algorithm to deflect probabilities on states modeling solutions. In the specific numerical experiments used for $T = 10$, the algorithm permits to have more than 62+11+11+12=94% of the distribution which is concentrated on states modeling solution (table 1).These results push us into considering that the deflection of the wave function complied with the theoretical considerations and permits to identify all the optimal solution in this specific example. Similar results have been obtained using both Qiskit and MyQLM library.

|  | Percentage T=10 | Percentage T=100 | Percentage T=1000 |
|---|---|---|---|
| $|111\rangle$ | 12.7% | 16.7% | 16.6% |
| $|100\rangle$ | 11.8% | 17.3% | 15.8% |
| $|101\rangle$ | 62.8% | 50.5% | 51.8% |
| $|001\rangle$ | 11.6% | 15.5% | 15.8% |

**Table 1. Adiabatic optimization with $T = 10, 100$ and $1000$**

The results prove that relatively small values of $T$ permits to deflect probabilities on solution states and larger $T$ parameters significantly improve the convergence.

The previous results were done on simulator and have been endorsed by experiments performed on the 27 qubits IBM quantum computer referred to as *ibmq_montreal* which is composed of 27 qubits and has an average CNOT error about $1.2.10^{-2}$ and an average readout error about $1.8.10^{-2}$ with gates time about 426ns. The Qiskit library is used to transpile and assign gates to the qubits (due to the fact that no triplet of qubits are directly fully entangled).

The results of table 2 prove the capacity of the ibmq_montreal to solve this 4-clauses and 3 variables SAT problem.

|  | Number of shots | percentage |
|---|---|---|
| $|111\rangle$ | 383 | 18.7% |
| $|100\rangle$ | 411 | 20.1% |
| $|101\rangle$ | 459 | 22.4% |
| $|001\rangle$ | 314 | 15.3% |
| $|010\rangle$ | 113 | 5.5% |
| $|000\rangle$ | 107 | 5.2% |
| $|110\rangle$ | 107 | 5.2% |
| $|011\rangle$ | 154 | 7.5% |

**Table 2. Adiabatic optimization with $T$=20 and 2048 shots**

82.4% of the distribution is on the 4 solutions of the SAT problem proving that the quantum computer provides a solution that meets the theory and that the average CNOT and average readout error permits to execute the circuit. However, the number of gates involved in such a circuit increases with $n$ and $T$ and large values force such circuits to be only executed on the simulator. Note that circuit quickly becomes cumbersome as the number of iterations increases with the number of gates that quickly grows, to permit execution on the current noisy quantum computers. Clearly low $T$ implementations are the most suitable for early quantum computers.

## 4. Concluding remarks

In this paper we give an explicit algorithm formalization of the adiabatic algorithm which can be considered as a new paradigm in a similar but quite different approach from the well-known simulated annealing.

The recent advances in quantum physics help us to define an efficient way to encode Boolean functions as the ground state of a Hamiltonian. The $Z$ gate and the $Z_i.Z_i...Z_k$ operators combined with the Hadfield's construction rules, can be used to explicitly define a problem mapping for Boolean function.

The programmable noisy intermediate-scale quantum computers now available are limited to the resolution of small SAT instances but the number of qubits in quantum computers has vastly increased during the last 5 years and noise probability has decreased.

The SAT resolution with adiabatic algorithm is a good pedagogical direction and can be seen as a first step into the design and analysis of quantum algorithms for more complex problems in Operation Research. Derived from quantum physics, adiabatic paradigm embeds problems into physical quantum hardware and permits to solve SAT and prove that applications of quantum theory should be of interest including complexity and eigenvalue problem computation. The class of simulated quantum annealing based approaches defined a new trend to solve the well-known difficulties in search space examination including diversification and local search intensification.